\documentclass[]{aa}  
\usepackage{natbib}

\usepackage{graphicx}
\begin{document}
%

\title{
Gamma-ray spectrum of RX~J1713.7$-$3946 in the {\it Fermi} era
and future detection of neutrinos
}


\titlerunning{Gamma-rays and Neutrinos from RX~J1713.7$-$3946}
\authorrunning{Yamazaki et al.}

\author{
Ryo Yamazaki
\inst{1},
Kazunori Kohri
\inst{2},
\and
Hideaki Katagiri
\inst{1}
}

\institute{
Department of Physical Science, Hiroshima University, 
Higashi-Hiroshima, Hiroshima 739-8526, Japan\\
\email{ryo@theo.phys.sci.hiroshima-u.ac.jp, katagiri@hep01.hepl.hiroshima-u.ac.jp}
\and
Physics Department, Lancaster University, LA1~4YB, UK\\
\email{k.kohri@lancaster.ac.uk}
}

   \date{Received August ??, 2008; accepted ??}

 
\abstract
{The recently launched  satellite,
\emph{Fermi Gamma-ray Space Telescope\/},
 is expected to 
find out if cosmic-ray (CR) protons are generated 
from supernova remnants (SNRs),
especially  RX~J1713.7$-$3946, by observing the GeV-to-TeV
$\gamma$-rays.
The GeV emission is thought to be
bright if the TeV emission is hadronic, i.e., of proton origin,
while dim if leptonic.
} 
{We reexamine the above view 
using a simple theoretical model of nonlinear acceleration of particles 
to calculate the gamma-ray spectrum of Galactic young SNRs.
}
{
}
{If the nonlinear effects of CR acceleration are
considered, it may be impossible to distinguish  the evidence of 
proton acceleration from leptonic  in the $\gamma$-ray spectrum
of Galactic young SNRs like RX~J1713.7$-$3946.
On the other hand, future  km$^3$-class neutrino observations will
likely find a clear evidence of the proton acceleration
there.
}
{}


\maketitle
%

\section{Introduction}

Recently, 
{\it Fermi Gamma-ray Space Telescope (Fermi)}\footnote
{http://fermi.gsfc.nasa.gov/}, 
observing GeV $\gamma$-ray photons, has been launched.
%
The GeV $\gamma$-ray observations with {\it Fermi}
are expected to identify the accelerators of Galactic 
cosmic-ray (CR) protons whose energy extends up to the
``knee'' energy ($\approx10^{15.5}$~eV).
%
At present, the most probable candidate for the
CR accelerator is a young supernova remnant (SNR).
Since the detections of
synchrotron X-rays  in some SNRs show 
 evidence of electron acceleration \citep{koyama95},
the current unsolved issue is whether the SNRs produce
high-energy protons or not.
TeV $\gamma$-ray observations are important to address this problem.
So far, TeV $\gamma$-rays  have been detected from several young SNRs
\citep{enomoto02,aharonian04,aharonian05c,katagiri05}.
They arise from either
leptonic (CMB photons up-scattered by high energy electrons) 
or hadronic ($\pi^0$-decay photons generated via accelerated
protons) processes, and it is generally difficult to separate 
these processes using  only  the TeV energy band;
 the study of wide-band, GeV-to-TeV spectra is necessary.

RX~J1713.7$-$3946 (hereafter RXJ1713) 
is a representative SNR from which 
bright TeV $\gamma$-rays have been detected.
The H.E.S.S. experiment measured 
the TeV spectrum and claimed that
 its shape was better explained by
the hadronic model \citep{hess06,hess07}.
So far, compared with other young SNRs, the TeV $\gamma$-ray 
spectrum of RXJ1713 is the most precisely measured and the 
energy coverage is wide, from 0.3 to 100 TeV, so that we can
obtain the best constraints on theoretical models.

Recently, time variation of synchrotron X-rays 
was discovered in RXJ1713 \citep{uchiyama07}.
If the variation timescale is determined from the
synchrotron cooling of X-ray emitting electrons,
the magnetic field is estimated to be $B\sim$~mG.
If so,  the leptonic, one-zone emission model 
\citep[e.g.,][]{aharonian99} cannot
explain the TeV-to-X-ray flux ratio, supporting
the hadronic origin of TeV $\gamma$-rays.
It should be noted that
the amplified magnetic field is theoretically expected
\citep[e.g.,][]{lucek00,giacalone07}.
In this case, according to the standard diffusive shock
acceleration theory,
the maximum energy of accelerated protons 
is estimated as \citep{aharonian99}
\begin{equation}
E_{{\rm max},p}=8\times10^3
\frac{B_{\rm mG}t_3}{\eta_g}
\left(\frac{v_s}{4000~{\rm km}~{\rm s}^{-1}}\right)^2
{\rm TeV}~~,
\end{equation}
which can be comparable to the knee energy.
Here, $B_{\rm mG}$, $v_s$, $t_3$, and $\eta_g$ are
the magnetic field strength in units of mG,
the shock velocity, the age of the SNR in units of
$10^3$~yr, and the gyrofactor, 
respectively.

However, at present, there are several issues to be addressed,
as the above picture on RXJ1713 is not yet proved.
First, if $B\sim{\rm mG}$ and TeV emission is hadronic, then in
order to explain the measured flux of radio synchrotron emitted by
primary electrons, the electron-to-proton ratio at the SNR should be
anomalously small, $K_{ep}\sim10^{-6}$ \citep{uchiyama03,butt08}, which
is far below the observed value at the earth and estimated values in
the nearby galaxy \citep{katz08}.
This might be resolved if the electrons are accelerated in the
later stages of SNR evolution, when the value of
$K_{ep}$ is different from the present value \citep{tanaka08},
although further discussions are necessary.
Second, the
hadronic scenario may be inconsistent with the 
molecular cloud (MC) observations \citep{fukui03}.
RXJ1713 is surrounded by MCs, which might suggest
 collision with them and high target number density.
If the TeV $\gamma$-rays are hadronic, 
such a region should be brighter than observed \citep{plaga07}.

In addition,
if the measured  width of the synchrotron X-ray  filaments at
the shock front of SNRs  is determined by the synchrotron cooling
effect \citep{uchiyama03,vink03,bamba03b,bamba05a,bamba05b}, the
magnetic field is independently
estimated as $B\approx0.1$~mG \citep{parizot06}, which is an order of
magnitude smaller than that estimated by \citet{uchiyama07}.
Also, the cutoff energy of TeV $\gamma$-ray spectrum is low,
so that in the one-zone hadronic scenario 
$E_{{\rm max},p}$ is estimated as 30--100~TeV \citep{villante07},
which is approximately two orders of
magnitude lower than the knee energy.
If $E_{{\rm max},p}<100$~TeV and $B\approx1$~mG, then
Eq.~(1) tells us $\eta_g\ga80$, implying far from
the ``Bohm limit'' ($\eta_g\approx1$) which is
inferred from the X-ray observation \citep{parizot06,yamazaki04}
or expected theoretically \citep{lucek00,giacalone07}.
This statement is recast if we involve  
recent results of X-ray observations.
The precise X-ray spectrum of RXJ1713 is revealed, which gives
$v_s=3.3\times10^8\eta_g^{1/2}$~cm~s$^{-1}$ \citep{tanaka08}.
Then, Eq.~(1) can be rewritten as
$E_{{\rm max},p}=5\times10^3B_{\rm mG}t_3$~TeV.
Hence, in order to obtain $E_{{\rm max},p}<100$~TeV,
we need $B\la20~\mu$G  in the context of the
hadronic scenario of TeV $\gamma$-rays.
One might think that the volume filling factor of the
region with $B\approx1$~mG is small and that the average
field strength is smaller, e.g., $B\approx0.1$~mG.
However, even in this case, $E_{{\rm max},p}$ is more than
100~TeV, which contradicts the observed $\gamma$-ray spectrum
beyond 10~TeV.

In these circumstances,
{\it Fermi} will give us important information on
 the $\gamma$-ray emission mechanism.
So far, the GeV emission has been thought to be
bright if the TeV emission is hadronic, while dim if leptonic.
However, this argument is not so straightforward
if the nonlinear model of CR acceleration is considered.
In the next section, we calculate the photon spectrum
using a simple semi-analytic model taking into account
 nonlinear effects.
Indeed, we show that in a certain case, the
hadronic emission spectrum in the GeV-to-TeV band
is  similar to the leptonic one.


\section{Hadronic gamma-rays in the efficient acceleration case }

If a large amount of protons are accelerated, their momentum flux is
large, so that the back-reaction of them is significant and the
background shock structure is modified
\citep{drury83,blandford87,malkov01}.  
Compared
with the test-particle (inefficient acceleration)  case in which the
back-reaction effects  are neglected, the background plasma is more
compressed at the shock due to the additional CR pressure, 
which leads to a harder CR spectrum.  
Hence the hadronic emission becomes harder\footnote{
Another kind of formation of a hard $\gamma$-ray
spectrum from accelerated protons is the SNR-MC interaction
system with appropriate separation
\citep[e.g.,][]{aharonian96,gabici07}.
The slower propagation of the low-energy protons
toward the cloud makes the  $\gamma$-ray spectrum hard.}.
At present, there is no reliable theory to determine the
acceleration efficiency, and it is not clear whether
this nonlinear model is correct or not.
Thus, the observations to determine the acceleration efficiency
and  the CR spectrum at the acceleration site are important.
It is widely expected that RXJ1713 with precise studies in
the $\gamma$-ray and X-ray bands is one of the best laboratories
to investigate theories of  nonlinear acceleration.

There are several models of nonlinear CR acceleration
\citep{berezhko94,ellison96,kang01,blasi02,blasi05,malkov97,amato05}.
Here, we adopt the one-dimensional, semi-analytic model 
\citep{blasi02,blasi05}. In the following,
we briefly summarize the formalism.
The accelerated CR protons are described by the
distribution function, $f(x,p)$, where $x$ is the
spatial coordinate and $p$ is the momentum of the accelerated 
proton.
We derive stationary solutions to
the set of an equation for $f(x,p)$ that describes the diffusive transport 
equation of accelerated protons,
and  equations for 
the background thermal plasma that is treated as a fluid.
The velocity, density, and thermodynamic properties of the fluid
can be determined by the mass and momentum conservation
equations, with the inclusion of the CR pressure  calculated as
\begin{equation}
P_{\rm CR}(x)=\frac{4\pi}{3}\int_{p_{\rm inj}}
^{p_{\rm max}}p^3v(p)f(x,p)\,dp~~.
\end{equation}
The injection of accelerated particles is assumed to occur
at the shock front ($x=0$), and mono-energetic
injection with the injection momentum $p_{\rm inj}$ is adopted.
This is because at present we do not understand the injection process
from first principles, hence we take a simple injection recipe.
We assume $p_{\rm inj}=\xi p_{\rm th}$,
where $p_{\rm th}=(2m_pk_BT_2)^{1/2}$ is
the momentum of particles in the thermal peak of the
Maxwellian distribution of the background plasma in the downstream region, 
having temperature $T_2$.
Furthermore, we assume
the continuity of the distribution function
of accelerated particles and the background thermal plasma
at $p=p_{\rm inj}$ and $x=0$, namely
$f(0,p_{\rm inj})=(n_2/\pi^{3/2}p_{\rm th}^3)
e^{-(p_{\rm inj}/p_{\rm th})^2}$, where $n_2$ is the
downstream number density of the background fluid.
In this formalism, input parameters are
the maximum momentum $p_{\rm max}$,
the upstream Mach number $M_0$, upstream fluid velocity $u_0$,
and the injection parameter $\xi$.
Given these parameters, 
the distribution function of accelerated protons
at the shock front $f_0(p)$ is calculated, and 
various physical quantities are obtained such as
the injection rate $\eta$,
the total compression ratio $R_{tot}$,
the fraction of the CR pressure at the shock
$\xi_c(0)=P_{\rm CR}(0) /\rho_0u_0^2$.

Note that recent nonlinear models of CR acceleration
have been developed taking into account the magnetic
field amplification \citep{vladimirov06} and 
its influence on turbulent heating \citep{vladimirov08}, 
fluid compression \citep{terasawa07,caprioli08a,caprioli08b}, 
and Alfv\'{e}nic drift \citep{zirakashvili08}, which are
neglected in the model considered in this paper.
These effects lead to less spectral hardening of
accelerated particles and smaller compression ratios,
and might be important in order to calculate the $\gamma$-ray
spectrum \citep{morlino08}. 
However, at present, 
it is not certain whether or not the magnetic field
is strongly amplified in the acceleration region;
although streaming instabilities between accelerated
protons and background plasma may occur, the nonlinear
evolution of the instability and the saturation level
are highly uncertain.
Although 
the magnetic field amplification is potentially
 coupled to the high injection rate of protons,
they should, in principle, be treated separately.
In this sense, our model  is the
extreme limit of the nonlinear acceleration theory,
which predicts the hardest spectrum of accelerated particles.

In this paper, we adopt $M_0=100$, 
$p_{\rm max}=1\times10^5m_pc$,
$u_0=5\times10^8$~cm~s$^{-1}$, and $\xi=3.6$.
Then, we obtain 
$p_{\rm inj}=3.74\times10^{-3}m_pc$,
$\eta=2.03\times10^{-4}$, $R_{tot}=36.2$, 
and $\xi_c(0)=0.902$.
While the total number of CR protons is much smaller
than that of the background plasma ($\eta\ll1$),
the CR pressure is dominant \citep{amato05}.
%
We find that the CR energy spectrum is asymptotically 
$N_p(p)\propto p^2f_0(p)\propto p^{-1.5}$, which is
harder than in the case of inefficient acceleration.
This result on the asymptotic form has been analytically
derived, which does not depend on the shock parameters
such as $M_0$ and $u_0$ in the large-$M_0$ limit
\citep{malkov97,malkov99}.

Using the derived distribution function of CR protons, 
we calculate the $\gamma$-ray spectrum
produced by $\pi^0$-decay process.
We used the PYTHIA Monte-Carlo event
generator \citep{Sjostrand:2006za}, which fits existing experimental
data well, to calculate the $pp$ scattering processes and detailed
distributions of the daughter particles such as $\pi^0$ and
$\pi^{\pm}$. We have also obtained the distribution functions of
emitted photons and neutrinos which are produced by subsequent decays
of those mesons and muons in the same code~\citep{yamazaki06}.
The result is shown in Fig.~1.

\begin{figure}
\includegraphics[width=20pc]{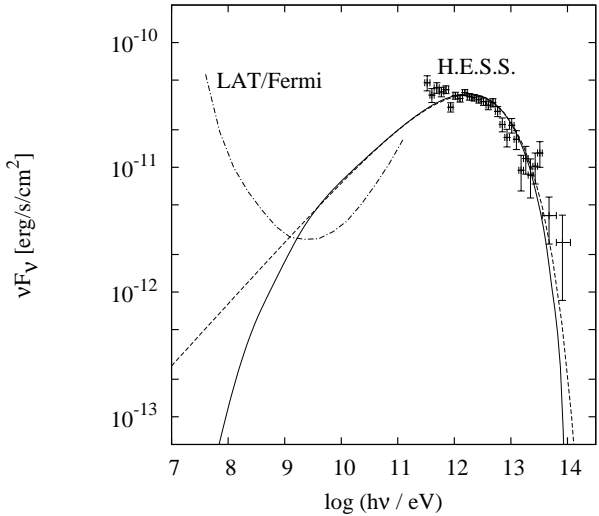}
\caption{
$\nu F_\nu$-spectra in $\gamma$-ray band, predicted by hadronic
$\pi^0$-decay model in the efficient acceleration  case (solid line)
and leptonic IC model in the inefficient case (dashed line).  The
dot-dashed curve shows the 1 year, 5 $\sigma$ sensitivity for the
Fermi LAT taking into account the Galactic diffuse background~
\citep{higashi08}.  The observed spectrum in the TeV band is shown
\citep{hess07}.  }
\end{figure}

So far, we have considered the hadronic $\gamma$-ray spectrum in the
context of the efficient acceleration scenario.
For comparison,  we show, in Fig.~1, the spectrum of
leptonic inverse-Compton (IC) radiation via accelerated electrons
in the case of inefficient acceleration, where back reaction effects of 
accelerated protons are neglected.
Throughout the paper, we take into account the Klein-Nishina
effect in calculating the IC spectrum.
The assumed form of the electron distribution is
$N_e(E_e)\propto E_e^{-s_e}\exp(-E_e/E_{\rm max})$,
and we adopt $s_e=2.0$ and $E_{\rm max}=28$~TeV.
This case can be realized if the magnetic field is weak enough
for the synchrotron cooling effect to be insufficient,
whose condition is written as
$t_{\rm age}<t_{\rm synch}=6\pi m_e^2c^3/\sigma_TE_{\rm max}B^2$,
where $t_{\rm age}=t_3\times10^3$~yr is the age of the SNR.
Solving this equation with $E_{\rm max}=28$~TeV, we derive
$B<26t_3^{-1/2}\mu$G.
If the magnetic field is strong ($B\gg26t_3^{-1/2}\mu$G), 
the spectral deformation
occurs, which will be discussed in \S~\ref{sec:conclusion}.
One can find from Fig.~1 that the 
$\pi^0$-decay $\gamma$-ray emission
in the efficient acceleration model coincides with 
the leptonic IC model in the inefficient case.
The reason is simple.
Let $s_i$ be the index of the energy spectrum of accelerated
particles $i$ ($i=p$ or $e$), so that
$N_i(E_i)\propto E_i^{-s_i}$.
Then, the radiation spectrum of $\pi^0$-decay $\gamma$-rays
is in the form $\nu F_\nu\propto\nu^{2-s_p}$, while the
spectrum of IC radiation is given by
$\nu F_\nu\propto\nu^{-(s_e-3)/2}$.
Hence, hadronic emission with $s_p\approx1.5$ and IC emission
with $s_e\approx2.0$ give the same spectral slope.
This is summarized in Table~\ref{table:1}.

Below several hundreds  of MeV,
hadronic $\gamma$-ray emission is dimmer than leptonic
 IC emission because $\pi^0$ creation reaction does not occur
 for low-energy ($<70$~MeV in the center-of-mass frame)
protons.
Unfortunately, {\it Fermi} sensitivity is not high enough to
recognize this decline below $\sim$~GeV.

One can find that both models slightly deviate from
the observed spectrum in the sub-TeV energy range.
The significance of this is sometimes strengthened,
because the leptonic
one-zone IC model is unlikely \citep{hess07}.
However, as will be seen in the next section,
it is not serious if the two-zone models are considered.

\begin{table*}
\caption{Spectral index of $\nu F_\nu$ spectrum of gamma-rays
($\nu F_\nu\propto\nu^\alpha$) for various cases.
Hadronic emission model in the case of efficient acceleration (Ia)
predicts similar $\gamma$-ray spectral slope with the leptonic,
inefficient acceleration model with weak magnetic field (IIIb).
On the other hand, hadronic inefficient acceleration models 
(IIa and IIIa) predict similar $\gamma$-ray spectral slope
with the leptonic, moderate magnetic field model (IIb).
}
\label{table:1}
\centering          
\begin{tabular}{c l | l | l}
\hline\hline       
&  & (a) $\pi^0$ model                
   & (b)  IC     model            \\
&  & ~~~~ $\alpha\approx2-s_p$ 
   & ~~~~ $\alpha=-(s_e-3)/2$ \\
\hline
I & Efficient acc. (strong $B$-field) & 
(Ia) $s_p\approx1.5$, $\alpha\approx0.5$   & 
(Ib) $s_e\approx2.5$, $\alpha\approx0.25$    \\  
\hline
II & Inefficient acc. (moderate $B$-field) & 
(IIa) $s_p\approx2$,   $\alpha\approx0$     & 
(IIb) $s_e\approx3$,   $\alpha\approx0$  \\  
\hline
III & Inefficient acc. (weak $B$-field) & 
(IIIa) $s_p\approx2$,   $\alpha\approx0$     & 
(IIIb) $s_e\approx2$,   $\alpha\approx0.5$  \\  
\hline                  
\end{tabular}
\end{table*}


\section{Two-zone models}

Here, we consider simple two-zone models to better explain 
the observed TeV spectrum \citep{aharonian99}.
RXJ1713 is interacting with MCs, so  that the environment around 
the shock producing high-energy particles may be inhomogeneous.
In this case, the one-zone approximation is too simplified, which
 motivates us to investigate the two-zone model
as the next-order approximation.
%

\subsection{Hadronic two-zone model}

In this model, two independent regions,
 $j$ ($j=1,2$), are considered.
For each component,
we independently calculate the proton spectrum again using 
the semi-analytic model of nonlinear CR
acceleration considered in the previous section.
The region $j$  has parameters
$M_0^{(j)}$, $p_{\rm max}{}^{(j)}$,
$u_0^{(j)}$, and $\xi^{(j)}$.
Then, we  derive the hadronic $\gamma$-ray spectrum
produced by $\pi^0$-decay process.
The total emission spectrum from the SNR
is simply given by the sum of
the emissions from two regions.

Figure~2 shows the result where we adopt
$p_{\rm max}{}^{(1)}=2\times10^4m_pc$ and
$p_{\rm max}{}^{(2)}=2\times10^5m_pc$.
The rest of the parameters are the same as those of the
previous section:
$M_0^{(1)}=M_0^{(2)}=100$, 
$u_0^{(1)}=u_0^{(2)}=5\times10^8$~cm~s$^{-1}$, 
and $\xi^{(1)}=\xi^{(2)}=3.6$.
%
%
The normalization of the $\pi^0$-decay emission 
is proportional to the product of the amount of the 
accelerated protons,
which is represented by $N_p(p=mc)$, 
and the target number density, $n_t$.
Here, we adjust 
$n_t^{(1)}N_p^{(1)}(p=mc)/n_t^{(2)}N_p^{(2)}(p=mc)=0.56$
in order to explain the observed $\gamma$-ray spectrum.
Then, one can see that the fit becomes better compared with
the one-zone hadronic model.

\begin{figure}
\includegraphics[width=20pc]{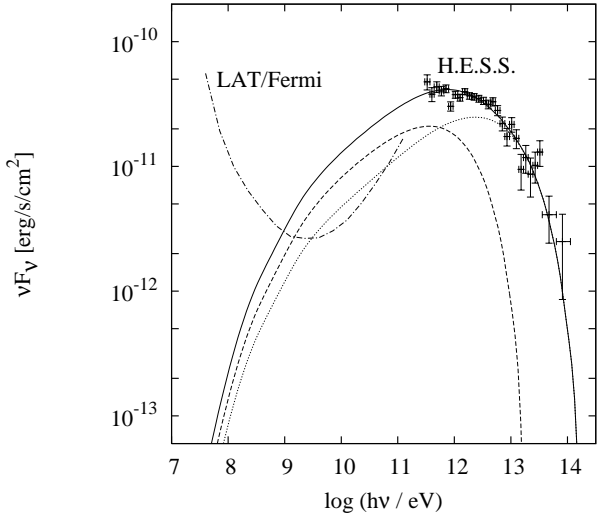}
\caption{
$\nu F_\nu$-spectra in the $\gamma$-ray band,
predicted by the two-zone hadronic model (solid line).
The dashed and dotted lines represent 
fluxes from region~1 and 2, respectively.
Others are the same as in Fig.~1.
}
\end{figure}

\subsection{Leptonic two-zone model}

The observed correlation between TeV $\gamma$-ray and synchrotron
X-rays \citep{hess06} may suggest that they have the same origin.
Then, since synchrotron X-rays arise from accelerated electrons,
one may expect that the leptonic model is likely
\citep{lazendic04,porter06,ogasawara07}.

Similar to the hadronic two-zone model,
two independent regions are considered.
The region $j$ ($j=1,2$) has a magnetic field $B^{(j)}$
and the electron spectrum 
$N^{(j)}(E_e)=A^{(j)}E_e^{-s_e^{(j)}}\exp(-E_e/E_{\rm max}^{(j)})$,
where $A^{(j)}$ is the normalization constant.
Here, the electron spectra are given by a single power-law form,
because the inefficient acceleration is adopted.
We consider synchrotron emission and IC emission in which
the target photon is the CMB.
The total emission spectrum from the SNR
is  given by the sum of
the emissions from two regions.

Figures 3 and 4 show the result
where we adopt $B^{(1)}=2.1~\mu$G, $B^{(2)}=10~\mu$G,
$s_e^{(1)}=s_e^{(2)}=2.0$, $E_{\rm max}^{(1)}=10$~TeV,
$E_{\rm max}^{(2)}=40$~TeV, and $A^{(1)}/A^{(2)}=2.74$.
The observed spectrum, including radio and X-ray bands,
can be explained by this model.
Note that
in the leptonic model,
the magnetic field strength must be much less than the
observationally inferred values \citep{uchiyama07,parizot06}
in order to fit the radio and X-ray synchrotron spectrum
---
if the magnetic field  were larger than 10~$\mu$G,
the predicted synchrotron radiation would be
much brighter than observed \citep{aharonian99}.
Hence, other explanations for
 the observations of rapid time variability
and thin width of synchrotron filaments
may be necessary \citep{pohl05,butt08,katz08}.
\begin{figure}
\includegraphics[width=20pc]{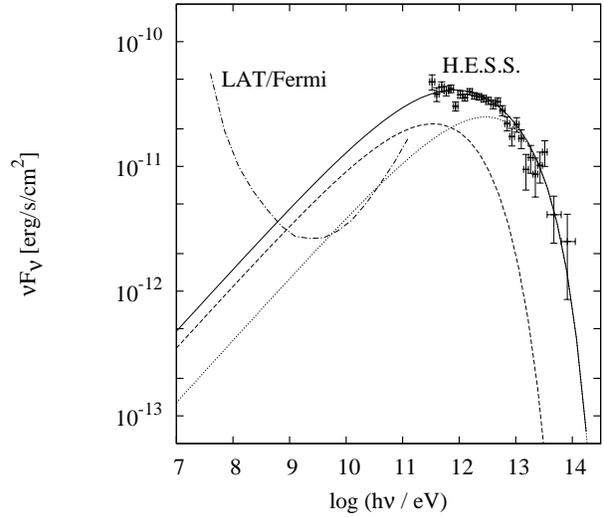}
\caption{
$\nu F_\nu$-spectra in the $\gamma$-ray band,
predicted by a two-zone leptonic IC model (solid line).
The dashed and dotted lines represent 
fluxes from region~1 and 2, respectively.
Others are the same as in Fig.~1.
}
\end{figure}
\begin{figure}
\includegraphics[width=20pc]{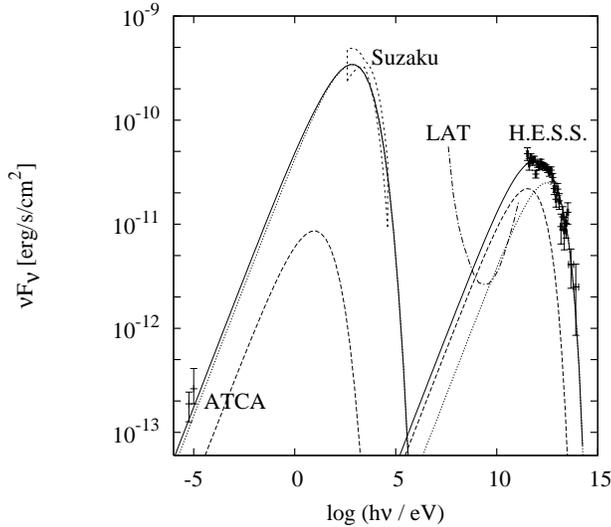}
\caption{
The same as in Fig.~3, but in the
wide-band energy range from radio to TeV $\gamma$-rays.
Radio and X-ray data are taken from \citet{hess06}
and \citet{takahashi08}, respectively.
}
\end{figure}

Comparing Fig.~2 with Fig.~3, we find that the
predicted spectrum of the hadronic two-zone model 
in the efficient acceleration case is
 similar to that of the leptonic two-zone model
in the case of inefficient acceleration.
This conclusion is the same as that for one-zone
models in \S~2.

\section{Conclusion}
\label{sec:conclusion}

Some  models predict relatively bright GeV $\gamma$-rays
compared with those considered above.
If the CR back-reaction effect on the particle spectrum
 is small (inefficient-acceleration case),
then the energy spectral index of protons is $s_p\approx2$, and
the $\pi^0$-decay $\gamma$-ray emission 
shows a roughly flat $\nu F_\nu$-spectrum, 
$\nu F_\nu\propto\nu^0$, in the GeV--TeV band
(model~IIa/IIIa in Table~\ref{table:1}).
The predicted flux is marginally consistent with the EGRET upper limit,
$\nu F_\nu\approx5\times10^{-11}$~erg~s$^{-1}$cm$^{-2}$ 
at 1--10~GeV \citep{hess06,hartman99}.
On the other hand,
if the magnetic field is moderately strong,
 the synchrotron cooling effect causes
steepening of the electron spectrum over a wide energy range 
--- typically $s_e\approx3$ 
\citep[e.g., see \S~19.3, eq.~(19.16) of][]{longair94}.
In this case, leptonic IC emission in the
GeV--TeV band again shows a nearly flat $\nu F_\nu$-spectrum
(model~IIb in Table~\ref{table:1}).
Therefore, these $\gamma$-ray 
emission models (IIa/IIIa and IIb) cannot
be distinguished.
This has been  discussed in \citet{ellison07},
where $B\approx60~\mu$G.

\begin{figure}
\includegraphics[width=20pc]{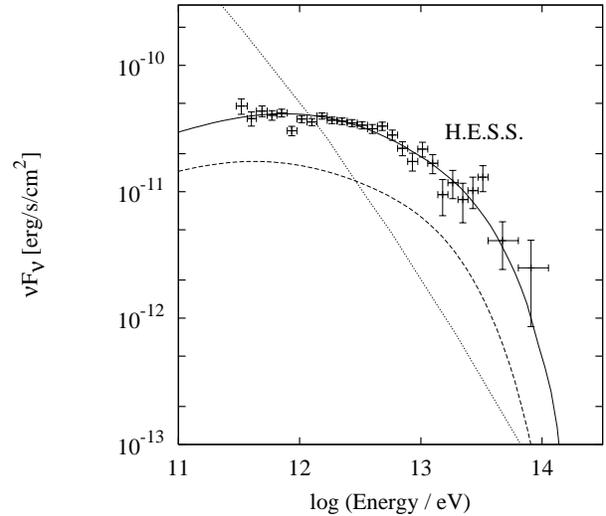}
\caption{
$\nu F_\nu$-spectra of TeV $\gamma$-rays (solid line) 
and $\mu$-neutrinos (dashed line)
calculated within the hadronic two-zone model in the efficient acceleration
case. The solid line is the same as in Fig.~2.
We have averaged the vacuum oscillation effects among neutrinos.
The dotted line shows the daily averaged atmospheric  neutrino
flux expected in KM3NeT~\citep{Kappes:2007ci,kappes07}.  }
\label{fig:neutrino}
\end{figure}

In summary,  it may be difficult to differentiate
between hadronic and leptonic  emission
by the spectral shape of the GeV-to-TeV $\gamma$-ray
emission of Galactic young SNRs like RXJ1713
(Table~\ref{table:1}).
As shown in this paper,
when the GeV $\gamma$-ray flux is  relatively low
(e.g., $\nu F_\nu\propto\nu^{0.5}$),
both an efficient acceleration model with hadronic $\gamma$-ray 
emission (model~Ia) and a leptonic, weak magnetic-field model
with inefficient acceleration (model~IIIb) may give 
similar spectral shapes.
On the other hand, as already pointed out in \citet{ellison07},
when the GeV emission is relatively bright
(e.g., $\nu F_\nu\propto\nu^0$),
one may not be able to distinguish 
the hadronic model in the inefficient case (models~IIa/IIIa)
from the leptonic one with a moderately strong magnetic
field  (model~IIb).
This conclusion may, at least qualitatively,  be applicable to
other young SNRs emitting TeV gamma-rays, such as 
RX~J0852.0$-$4622 \citep{katagiri05,aharonian05c}.
{\it Fermi} will likely provide us with
rich information on the emission mechanism of RXJ1713 and other
young SNRs.
However, one should only draw  conclusions with great care,
even in the {\it Fermi} era. 
Probably,   neutrino observation with km$^3$-class detectors such as
IceCube~\citep{Achterberg:2007qp} or KM3NeT~\citep{Kappes:2007ci}
will finally resolve  the problem
\citep{crocker02,halzen02,kistler06,vissani08,huang08,halzen08}.
As shown in Fig.~\ref{fig:neutrino}, if the observed TeV $\gamma$-ray
emission is hadronic, then  the expected neutrino spectrum at the
source is above the atmospheric neutrino background at around 5 --
10~TeV, which may become `the smoking gun' of proton acceleration
in Galactic young SNRs.

\begin{acknowledgements}
We are grateful to the referee for useful comments.
This work was supported in part by a Grant-in-aid from the 
Ministry of Education, Culture, Sports, Science,
and Technology (MEXT) of Japan,
No.~18740153, No.~19047004 (R.~Y.) and No.19740143 (H. K.), and
in part by  PPARC grant, 
PP/D000394/1, EU grant MRTN-CT-2006-035863, 
the European Union through the Marie Curie Research and 
Training Network ``UniverseNet'',
MRTN-CT-2006-035863 (K. K.)
\end{acknowledgements}


\bibliographystyle{aa}

\end{document}